\documentclass[prl,showpacs,twocolumn, amsmath,amssymb,superscriptaddress,floatfix,nofootinbib]{revtex4-1}
\usepackage[utf8]{inputenc}
\usepackage{textcomp}
\usepackage[sort&compress]{natbib}
\usepackage[normalem]{ulem}
\usepackage{lipsum} % 调用跨栏长公式宏包
\usepackage{bm}
\usepackage{times}
\usepackage{amssymb,amsbsy,amsmath,amsfonts}
\usepackage{graphicx}
\usepackage{float}
\usepackage{caption}
\usepackage{wrapfig}
\usepackage{subcaption}
\usepackage{color}
\usepackage{morefloats}
\usepackage{rotating}
\usepackage{srcltx}
\usepackage{slashed}
\usepackage{multirow}
\usepackage{verbatim}
\usepackage{hyperref}
\usepackage{tabularx}
\usepackage{adjustbox}
\usepackage[dvipsnames]{xcolor}
\usepackage{nicefrac}
\usepackage{siunitx}
\renewcommand{\arraystretch}{1.2}
\usepackage{hyperref}
\hypersetup{
	colorlinks	=true,
	urlcolor	=blue,
	linkcolor	=blue,
	citecolor	=blue,
	pdftitle	={},
	pdfauthor	={},
	pdfsubject	={femtoscopy}
	}


\begin{document}

\title{Revealing the nature of double-strangeness pentaquark states via femtoscopic correlation functions}

\author{Zheng-Ting Lai}
\affiliation{School of Physics, Beihang University, Beijing 102206, China}

\author{Jun-Xu Lu}
\email[Corresponding author: ]{ljxwohool@buaa.edu.cn}
\affiliation{School of Physics, Beihang University, Beijing 102206, China}

\author{Zhi-Wei Liu}
\email[Corresponding author: ]{liuzhw@szu.edu.cn}
\affiliation{Institute for Advanced Study in Nuclear Energy \& Safety, College of Physics and Optoelectronic Engineering, Shenzhen University, Shenzhen 518060, China}
\affiliation{Shenzhen Key Laboratory of Nuclear and Radiation Safety, Shenzhen 518060, China}

\author{A. Martínez Torres}
\email[Corresponding author: ]{amartine@if.usp.br}
\affiliation{Universidade de Sao Paulo, Instituto de Física, C.P. 05389-970, Sao Paulo, Brazil}

\author{K. P. Khemchandani}
\email[Corresponding author:]{
kanchan.khemchandani@unifesp.br}
 \affiliation{Universidade Federal de São Paulo, C.P. 01302-907, São Paulo, Brazil}

\author{Li-Sheng Geng}
\email[Corresponding author: ]{lisheng.geng@buaa.edu.cn}
\affiliation{School of Physics, Beihang University, Beijing 102206, China}
\affiliation{Peng Huanwu Collaborative Center for Research and Education, Beihang University, Beijing 100191, China}
\affiliation{Southern Center for Nuclear-Science Theory (SCNT), Institute of Modern Physics, Chinese Academy of Sciences, Huizhou 516000, China}

\begin{abstract}
Recent discoveries of exotic hadrons, which cannot be classified within the conventional quark model of $q \bar{q}$ mesons and $qqq$ baryons, strongly imply the existence of dynamically generated hadronic molecules. Some of these hadron-hadron interactions are accompanied by coupled-channel effects, which remain challenging to quantitatively determine. In this work, we demonstrate that femtoscopy provides a sensitive probe of such coupled-channel dynamics. We calculate correlation functions for the double-strangeness pentaquark candidates $P_{css}(4493)$ ($J^P=1/2^-$) and $P_{css}(4633)$ ($J^P=1/2^-$ or $3/2^-$), revealing clear signatures of the attractive interactions that can form bound states. The results are markedly different from those obtained in scenarios that neglect off-diagonal transitions, highlighting the importance of coupled-channel effects for understanding the structure of these hadrons.
\end{abstract}

%\pacs{13.75.Gx, 13.75.Jz,12.39.Fe}

\maketitle

%%%%%%%%%%%%%%%%%%%%%%%%%%%%%%%%%
%\section{Introduction}
%%%%%%%%%%%%%%%%%%%%%%%%%%%%%%%%%

\section{Introduction}

Quantum Chromodynamics (QCD), the fundamental theory governing the strong interaction, dictates that quarks are confined within color-singlet states, i.e., hadrons, in the low-energy region. Consequently, understanding hadron-hadron interactions is a core goal of modern theoretical particle and nuclear physics, as it can deepen our understanding of the non-perturbative behavior of QCD~\cite{ALICE:2020mfd,Fabbietti:2020bfg} and is essential for studies of beyond-the-Standard-Model physics~\cite{Brambilla:2014jmp,Belley:2023lec,Aliberti:2025beg}.

Since 2003, numerous exotic hadrons with structures beyond the conventional $q\bar{q}$ mesons and $qqq$ baryons have been observed~\cite{Oset:2016lyh, Richard:2016eis,Hosaka:2016ypm,Chen:2016qju,Esposito:2016noz,Lebed:2016hpi,Guo:2017jvc,Ali:2017jda,Olsen:2017bmm,Karliner:2017qhf,Liu:2019zoy,vanBeveren:2020eis,Chen:2022asf,Mai:2022eur,Bai:2026atm}. Various interpretations have been proposed, including compact multi-quark states, hadronic molecules, glueballs, and quark-gluon hybrid states. Some of these observed exotic structures may alternatively arise from kinematic effects, such as triangle singularities~\cite{Liu:2015fea,Xie:2016lvs,Guo:2019twa,COMPASS:2020yhb}. Among these interpretations, hadronic molecules are the most favored because most of these hadrons lie near the two-hadron thresholds of conventional hadron pairs~\cite{Guo:2017jvc,Liu:2024uxn}. This molecular picture is analogous to the deuteron, a bound state of a proton and a neutron~\cite{Weinberg:1965zz}. It is worth mentioning that research on exotic three-hadron molecular states has also advanced significantly~\cite{Ren:2018pcd,MartinezTorres:2018zbl,Wu:2022ftm}.

The interpretation of hadronic molecules relies heavily on our understanding of hadron-hadron interactions. However, accurate characterization of these hadron-hadron interactions is severely hampered by the scarcity of experimental data, with only a few exceptions, such as nucleon-nucleon and $\pi N$ interactions. The primary reason is that these hadrons, typically hyperons or heavy-flavored hadrons, have extremely short lifetimes. It is impossible to produce them in sufficient amounts to form fixed targets or particle beams, thereby making conventional scattering experiments infeasible. Therefore, hadron-hadron interactions associated with hadronic-molecule candidates are mainly constrained by invariant-mass distributions. Recently, three approaches have been reviewed in Ref.~\cite{Liu:2024uxn} to help reveal the nature of hadronic molecules, including studying multiplets of hadronic molecules based on heavy-quark spin/flavor symmetry or SU(3) symmetry, studying three-body hadronic molecules that can help verify the molecular nature of related two-body systems, and employing the femtoscopic technique with which one can extract insights into hadron-hadron interactions directly from $pp$, $pA$, and $AA$ collisions. Notably, the femtoscopic technique can reveal the internal structure of many such exotic hadrons~\cite{Liu:2023wfo,Liu:2024nac,Liu:2025nze,Liu:2025wwx,Vidana:2023olz,Abreu:2024qqo}.

In the present work, we focus on a key issue in hadron-hadron interactions: coupled-channel effects, i.e., off-diagonal transitions. Quantum mechanics allows the mixing of different reaction channels with identical quantum numbers, a manifestation of underlying symmetries. Thus, in addition to elastic channels, scattering processes naturally involve inelastic channels. Coupled-channel effects modify the scattering amplitudes and consequently alter the pole structure of the scattering matrix, ultimately resulting in observable signatures in invariant-mass distributions, including threshold effects and line shapes. The well-known two-pole structure of the $\Lambda(1405)$~\cite{Oller:2000fj,Jido:2003cb,Meissner:2020khl,Xie:2023cej} is a typical manifestation of coupled-channel effects. In scattering processes featuring identical initial and final hadron pairs, coupled channels merely emerge as intermediate states, whose contributions are frequently treated as minor effects, though this assumption is not always justified. 

Inspired by the capability of femtoscopy to extract interaction information and reveal the intrinsic nature of hadronic states, we further demonstrate in this work that the technique is sensitive to coupled-channel dynamics and can uncover the essential properties of related hadronic molecules. We take the double-strangeness pentaquark states $P_{css}$, dynamically generated via charmed meson-baryon interactions, as typical examples in which coupled-channel effects dominate.

The existence of double strangeness hidden-charm pentaquark states has been expected since the discovery of the pentaquark states $P_c(4380)$, $P_c(4312)$, $P_c(4440)$, $P_c(4457)$, $P_c(4337)$~\cite{LHCb:2015yax,LHCb:2019kea,LHCb:2021chn} and strange hidden-charm pentaquark states $P_{cs}(4459)$~\cite{LHCb:2020jpq} and $P_{cs}(4338)$~\cite{LHCb:2022ogu}, though no solid experimental evidence has been found so far. At present, the pentaquark states are widely interpreted as meson-baryon molecular states, though several ambiguities remain, such as the spins of $P_c(4440)$ and $P_c(4457)$. See Refs.~\cite{Chen:2016qju,Lebed:2016hpi,Esposito:2016noz,Guo:2017jvc,Ali:2017jda,Liu:2019zoy,Chen:2022asf,Liu:2024uxn} for recent reviews. Double-strangeness hidden-charm pentaquark states were first proposed in Ref.~\cite{Wang:2020bjt} within the light-meson (no pion) exchange model, but this approach requires an unphysical regularization cutoff. Later in Ref.~\cite{Marse-Valera:2022khy}, via the vector meson exchange model, two $P_{css}$ states of molecular nature located at $(M_R,\Gamma_R)=(4493.35, 73.67)$ MeV due to pseudoscalar meson-baryon (PB) interactions, denoted as $P_{css}(4493)$ with $J^P=1/2^-$, and $(4633.38, 79.58)$ MeV due to vector meson-baryon (VB) interactions, denoted as $P_{css}(4633)$ with $J^P=1/2^-$ or $3/2^-$, are predicted. The two bound states are generated predominantly through coupled-channel dynamics. The dominant attraction arises from the strong off-diagonal interactions between the two heaviest meson-baryon channels, whereas the diagonal interactions are significantly suppressed or even repulsive. A similar model was applied in Ref.~\cite{Roca:2024nsi}, using an alternative approach to evaluate the $BBV$ verticesleading to the prediction of four $P_{css}$ states.

In the present work, by taking the $P_{css}(4493)$ and $P_{css}(4633)$ states dominated by coupled-channel effects as representative examples, we investigate how coupled-channel dynamics is reflected in femtoscopic momentum correlation functions. The manuscript is organized as follows. In Sec.~II, we briefly revisit the mechanisms underlying the dynamical generation of the $P_{css}$ states and introduce the Koonin-Pratt (KP) formalism for correlation functions. In Sec.~III, we present the numerical results and corresponding discussions. Finally, a conclusion is drawn in Section IV.

\section{Formalism}

In this section, we briefly review the mechanisms underlying the dynamical generation of the two $P_{css}$ states in the unitarized coupled-channel approach based on vector-meson exchange interactions. More details are available in Ref.~\cite{Marse-Valera:2022khy}. 
\begin{table}[h]
\centering
\caption{Thresholds of the coupled channels considered in the PB and VB systems.}
\label{tab:PB_VB_masses}
\begin{tabular}{l c l c}
\hline\hline
\multicolumn{2}{c}{PB System} & \multicolumn{2}{c}{VB System} \\
\hline
Channel & Threshold (MeV) & Channel & Threshold (MeV) \\
\hline
$\eta_c \Xi^-$ & 4298 & $J/\psi \Xi^-$ & 4415 \\
$D_s^- \Xi_c^0$ & 4437 & $D_s^{*-} \Xi_c^0$ & 4581 \\
$D_s^- \Xi_c^{\prime 0}$ & 4545 & $D_s^{*-} \Xi_c^{\prime 0}$ & 4689 \\
$D^- \Omega_c^0$ & 4564 & $D^{*-} \Omega_c^0$ & 4706 \\
\hline\hline
\end{tabular}
\end{table}
The thresholds for the relevant channels are listed in Table~\ref{tab:PB_VB_masses}. Notably, these channels are not subject to the Coulomb interaction, which allows the contributions from the strong interaction to be isolated.

\subsection{Dynamical generation of the $P_{css}(4493)$ and $P_{css}(4633)$}

Following Ref.~\cite{Marse-Valera:2022khy}, when considering the interactions between the pseudoscalar mesons and baryons, the coupled channels in the particle basis are $\eta_{c} \Xi^-$, $D_s^- \Xi_{c}^0$, $D_s^- \Xi_{c}^{\prime0}$, and $D^- \Omega_{c}^0$. Note that such a choice of charge states ensures that the interaction is free of Coulomb effects. The \textit{t}-channel vector-meson exchange diagrams reduce to contact interactions up to $\mathcal{O}(p^2)$, where $p$ denotes a generic small momentum. Upon projection onto the $S$-wave, the potential becomes
\begin{align}
V_{ij}(\sqrt{s})=-C_{ij} \frac{1}{4 f^{2}}\left(2 \sqrt{s}-M_{i}-M_{j}\right) N_{i} N_{j}
\label{eq:ponPcss}
\end{align}
where $N_{i}$ and $N_{j}$ are the normalization factors for baryon spinors, $N_{i}= \sqrt{\left(E_{i}+M_{i}\right) / 2 M_{i}}$, and  $M_{i}\left(E_{i}\right)$ and $M_{j}\left(E_{j}\right)$ are the masses (energies) of the baryons in the incoming and outgoing channels, respectively. The $C_{ij}$ coefficients of the PB interaction are listed in Table~\ref{tab:Cij_PB}. 

\begin{table}[htbp]
\centering
\caption{ $C_{ij}$ coefficients of the $PB$ interaction in the $(I,S)=(1/2,-2)$ sector, where $\kappa_{c} \simeq 1 / 4$ and $\kappa_{c c} \simeq 1 / 9$.}
\label{tab:Cij_PB}
\renewcommand{\arraystretch}{1.3}
\begin{tabular}{c|cccc}
\hline\hline
 & $\eta_{c} \Xi^-$ 
 & $D_s^- \Xi_{c}^0$ 
 & $D_s^- \Xi_{c}^{\prime0}$ 
 & $D^- \Omega_{c}^0$ \\
\hline
$\eta_{c} \Xi^-$
 & $0$
 & $\sqrt{\frac{3}{2}}\kappa_c$
 & $\frac{1}{\sqrt{2}}\kappa_c$
 & $-\kappa_c$ \\

$D_s^- \Xi_{c}^0$
 & 
 & $-1+\kappa_{cc}$
 & $0$
 & $0$ \\

$D_s^- \Xi_{c}^{\prime0}$
 & 
 & 
 & $-1+\kappa_{cc}$
 & $-\sqrt{2}$ \\

$D^- \Omega_{c}^0$
 & 
 & 
 & 
 & $\kappa_{cc}$ \\
\hline\hline
\end{tabular}
\end{table}

We employ the Bethe-Salpeter equation in coupled channels to account for the non-perturbative effects,
\begin{align}\label{eq:BSeq}
T_{ij}=V_{ij}+V_{il} G_{l} T_{lj},
\end{align}
where $G_l$ denotes the meson and baryon propagator labeled by the $l$-th intermediate channel. Under the on-shell approximation, the above integral equation reduces to a simple algebraic one: $T=(1-V G)^{-1}V$, where $G$ is a diagonal matrix whose elements are meson-baryon loop functions defined as
\begin{align}
G_{l}(P)=i \int \frac{d^{4} q}{(2 \pi)^{4}} \frac{2 M_{l}}{(P-q)^{2}-M_{l}^{2}+i \epsilon} \frac{1}{q^{2}-m_{l}^{2}+i \epsilon},\label{Gl}
\end{align}
in which $M_{l}$ and $m_{l}$ correspond to the masses of baryon and meson inside the loop. In Eq.~(\ref{Gl}), $P=p+k=(\sqrt{s},\bf{0})$ is the total four-momentum in the center of mass frame, and $q$ is the internal meson four-momentum.

We closely follow the treatment of Ref.~\cite{Marse-Valera:2022khy} and employ the \textit{cutoff method} with $\Lambda=800$ MeV to regularize the loop function $G_l$. A double-strangeness pentaquark state with a mass of $M_R=4493$ MeV and a width of $\Gamma_R=74$ MeV is dynamically generated and strongly couples to $D^- \Omega_{c}^0$. As mentioned in Ref.~\cite{Marse-Valera:2022khy}, the attractive interaction of the $D^- \Omega_{c}^0$ in the channel alone is insufficient to generate the state, and the off-diagonal transition amplitude $D^- \Omega_{c}^0\rightarrow D_s^- \Xi_{c}^{\prime0}$ provides the additional attraction required. A similar picture is also found in Ref.~\cite{Roca:2024nsi}, where the $VVB$ vertices are evaluated by directly using the flavor and spin wave functions after appropriate symmetrization~\cite{Debastiani:2017ewu, Wang:2022aga}, instead of employing the SU(4) symmetry adopted in Ref.~\cite{Marse-Valera:2022khy}.

On the other hand, in Ref.~\cite{Clymton:2025zer}, an off-shell coupled-channel formalism based on the one-boson-exchange model was applied, and a double-strangeness pentaquark at $(M_R,\Gamma_R)=(4504,0.4)$ MeV was found. The three dominant coupled channels are $D_s^- \Xi_{c}^0$, $D_s^- \Xi_{c}^{\prime0}$, and $D^- \Omega_{c}^0$. However, the $P_{css}$ state was shown to persist even without the $\bar{D}_s \Xi_{c}^{\prime}$ contribution~\cite{Clymton:2025zer}, indicating a distinct binding mechanism compared with the one in Ref.~\cite{Marse-Valera:2022khy}.

The $P_{css}(4633)$ state in the vector-meson-baryon sector is generated analogously. In the VB sector, four coupled channels in the particle basis including $J/\psi \Xi^-$, $D_s^{\ast-} \Xi_c^0$, $D_s^{\ast-} \Xi_c^{\prime0}$, and $D^{\ast-} \Omega_c^0$ are taken into account, in which the $P_{css}(4633)$ state is dominated by the transition amplitude between the last two heavier channels. Note that the VB interaction has the same structure and coupling strength as the PB interaction, except for the scalar product between the polarization vectors of the incoming and outgoing vector mesons.

\subsection{Femtoscopic correlation functions}

Femtoscopic correlation functions are calculated via the Koonin-Pratt (KP) formula~\cite{Koonin:1977fh,Pratt:1990zq},
\begin{align}
   C(\mathbf{p}_1,\mathbf{p}_2)&\simeq \int d \mathbf{r} S_{12}(r)|\Psi(\mathbf{r},\mathbf{k})|^2 ,\label{Corr}
\end{align}
where we adopt a Gaussian source function \begin{align}
S_{12}(r) &= \frac{1}{(2 \sqrt{\pi} R)^3} \, \text{e}^{-r^2/(4 R^2)}.
\end{align}

In Eq.~(\ref{Corr}), the $\Psi(\mathbf{r},\mathbf{k})$ is the relative wave function of the two-body outgoing state, which, taking the on-shell approximation, can be written as~\cite{Vidana:2023olz}:
\begin{align}
   \Psi(\mathbf{r},\mathbf{k})=e^{i\mathbf{k}\cdot\mathbf{r}} +T(k,k)\cdot\theta(q_{\rm{max}}-k)\cdot \tilde{G}(r,\sqrt{s}),
\end{align}
where $T(k,k)$ is the scattering amplitude in Eq.~(\ref{eq:BSeq}) and $\tilde{G}(r,\sqrt{s})$ is given by
\begin{align}
   \tilde{G}(r,\sqrt{s})=&\int_0^{q_{\rm{max}}} \frac{d^3\mathbf{k'}}{(2\pi)^3}\frac{ M_B}{E_B(\mathbf{k'})E_M(\mathbf{k'})} \nonumber\\
   &\cdot\frac{j_0(k^\prime r)}{\sqrt{s}-(E_B(\mathbf{k'})+E_M(\mathbf{k'}))+i\epsilon},
\end{align}
where $E_{B}(\mathbf{k'})=\sqrt{M_B^2+\mathbf{k'}^2}$ and $E_{M}(\mathbf{k'})=\sqrt{M_M^2+\mathbf{k'}^2}$ are the energies of the corresponding baryon and meson. The $q_{\rm{max}}$ are set at $\Lambda=800$ MeV in the calculation of $G_l$.

Substituting the wave function into the KP formula and considering coupled channels, one obtains 
\begin{align}\label{eq:CFs}
   C_i(k)=&1+\theta(q_{\rm{max}}-k)\int_0^\infty d r~4\pi r^2S_{12}(r) \nonumber \\
   &\times\Bigg[ |j_0(kr)+T_{ii}(\sqrt{s})\cdot \tilde{G}_i(r,\sqrt{s})|^2 - |j_0(kr)|^2  \nonumber\\
   &+ \sum_{i\ne j}\omega_j|T_{ji}(\sqrt{s})\cdot\tilde{G}_j(r,\sqrt{s})|^2\Bigg],
\end{align}
where $\omega_j$ is the weight for each component of the multichannel wave function, and the sum runs over all possible coupled channels. For simplicity, we assume the weights are all equal to 1 in this study. The subscript $i,j$ label the corresponding coupled channels and $\sqrt{s}=E_B(k)+E_M(k)$.

\begin{figure*}[htbp]
    \centering
    \includegraphics[width=0.55\textwidth]{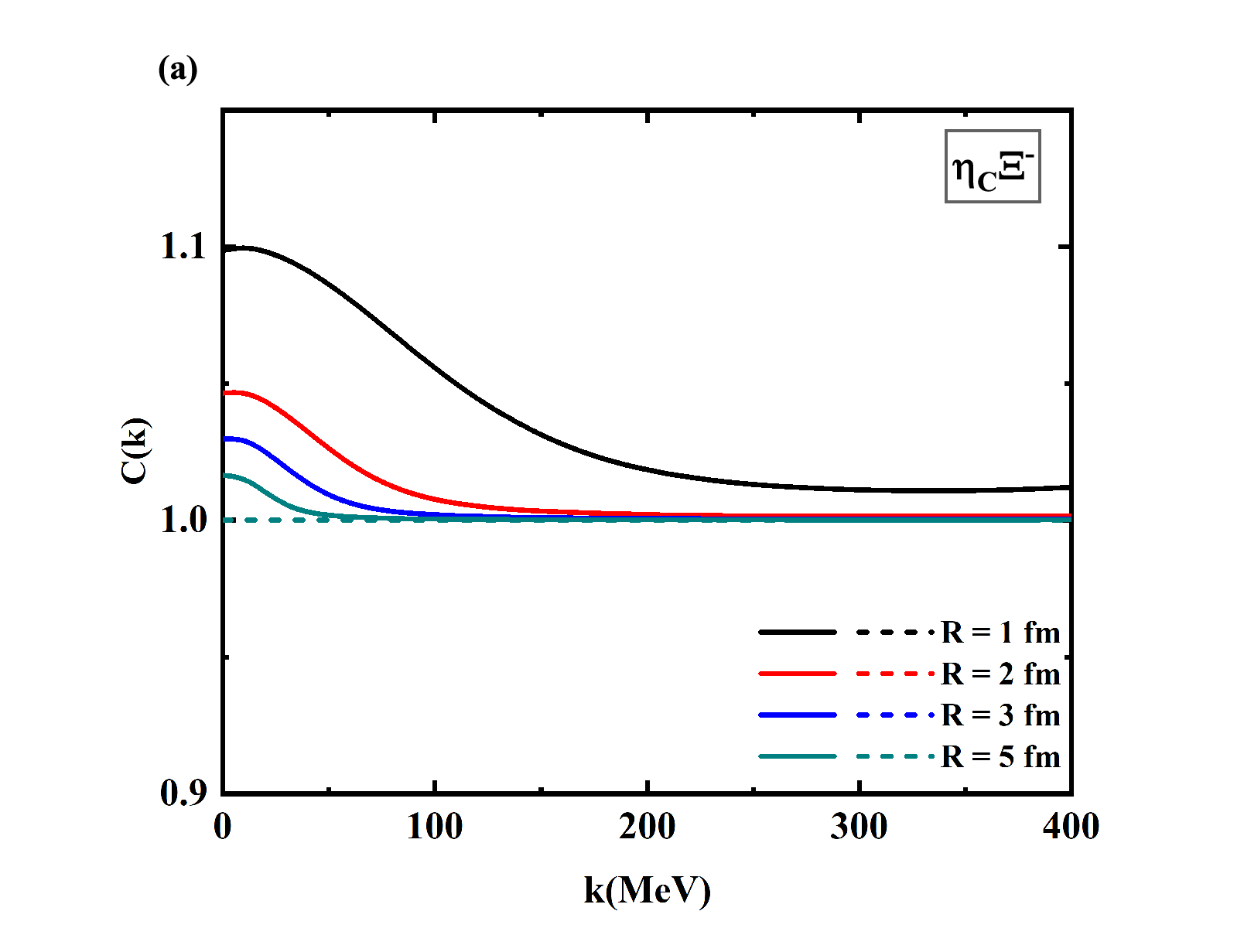}%
    \hspace{-0.1\textwidth}%
    \includegraphics[width=0.55\textwidth]{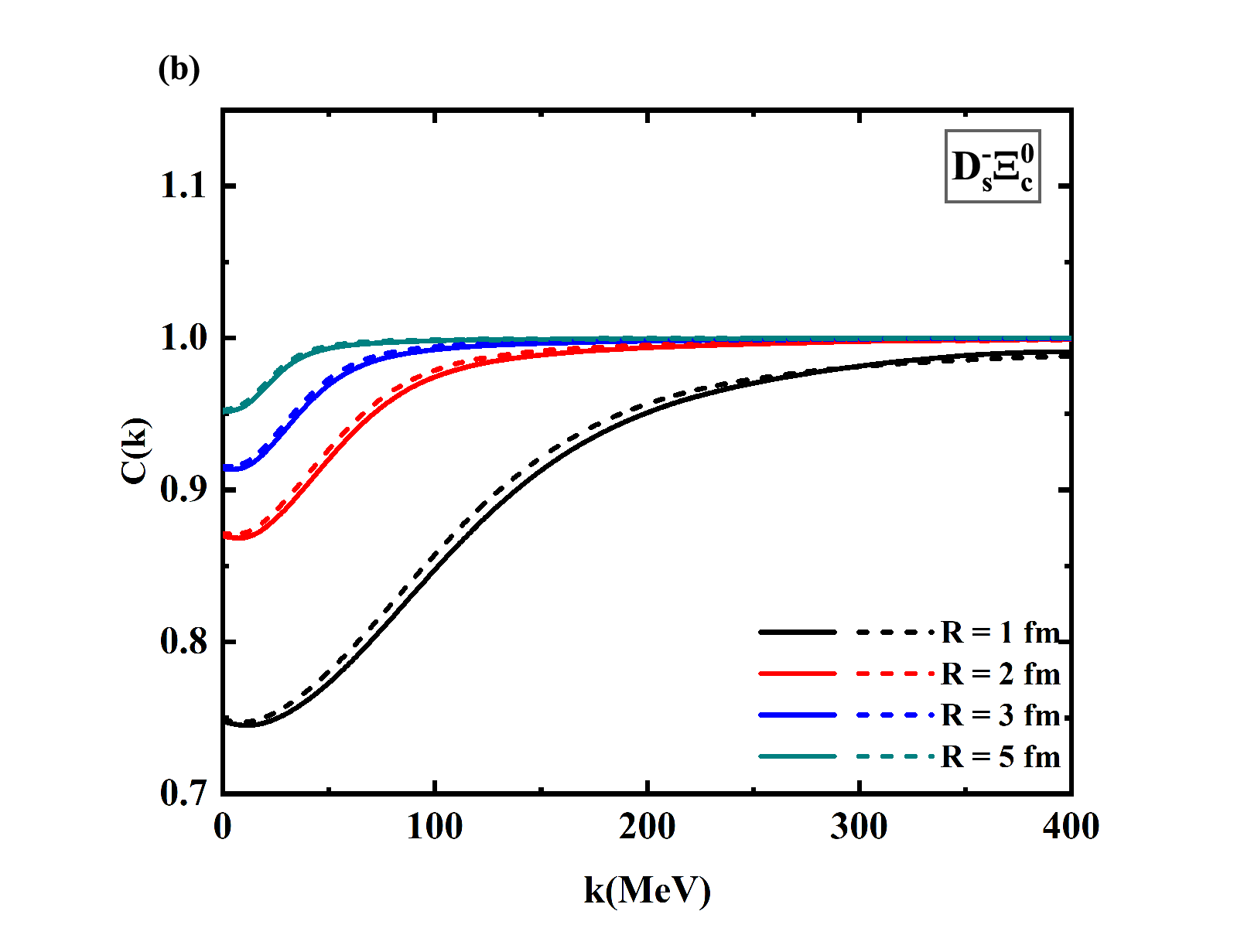}\\[-25pt]
    \includegraphics[width=0.55\textwidth]{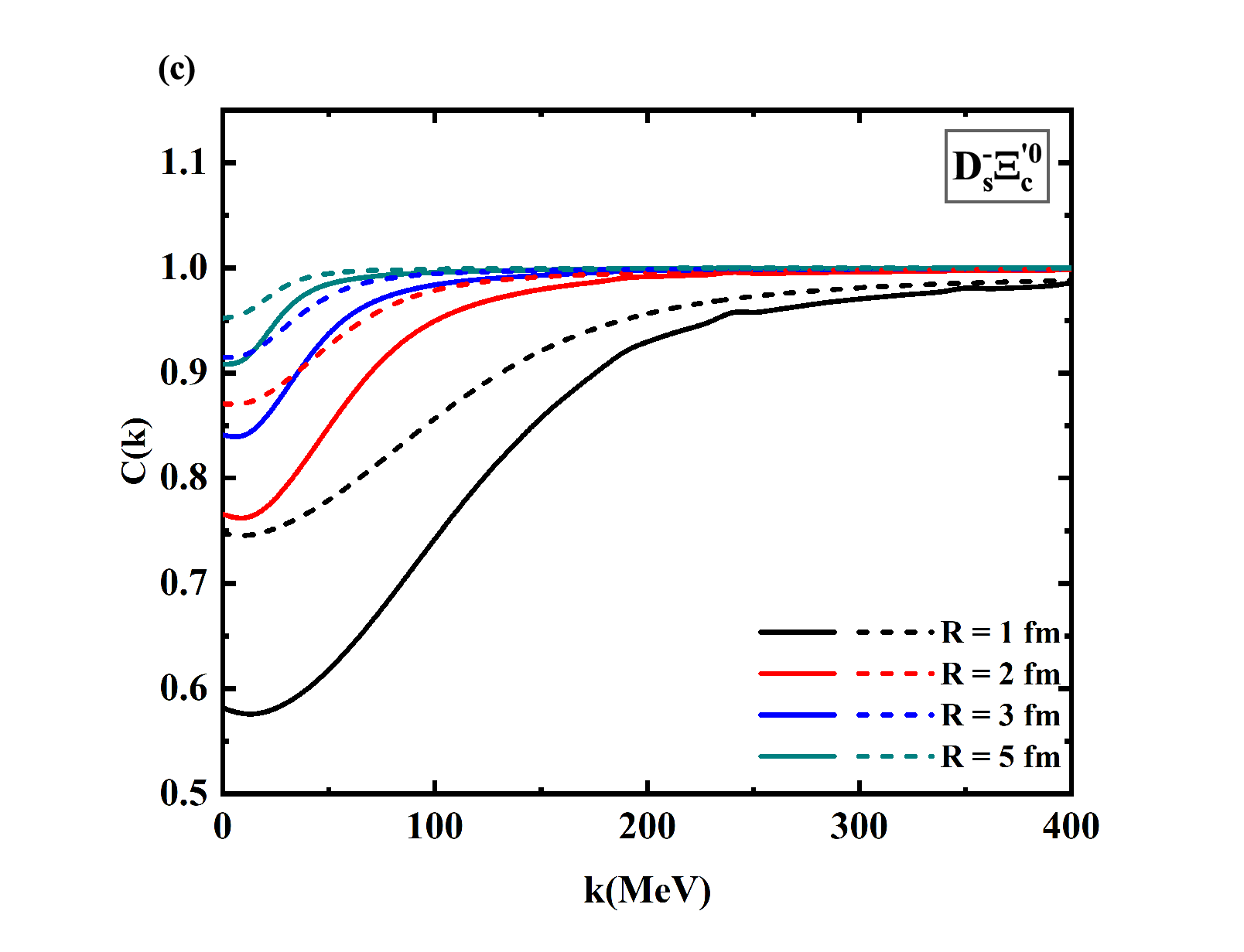}%
    \hspace{-0.1\textwidth}%
    \includegraphics[width=0.55\textwidth]{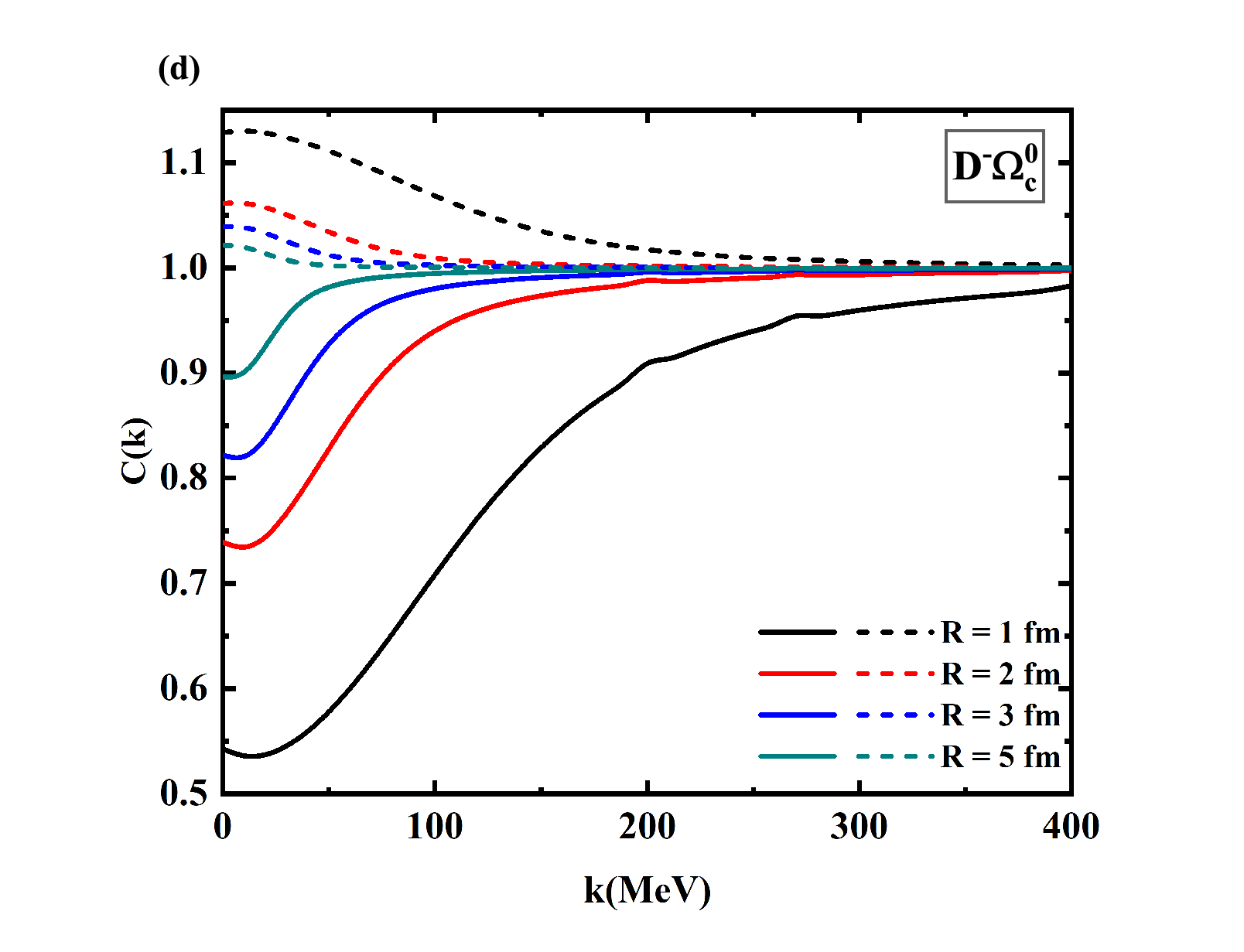}
    \caption{$\eta_{c} \Xi^-$ (a), $D_s^- \Xi_{c}^0$ (b), $D_s^- \Xi_{c}^{\prime0}$ (c), and $D^- \Omega_{c}^0$ (d) correlation functions as a function of the relative momentum $k$ for different source sizes $R=1$, 2, 3, and 5 fm. The solid lines represent the results obtained in the coupled-channel case, while the short dashed lines represent the results obtained in the single-channel scattering process.}
    \label{CF_pcss_PB}
\end{figure*}

\begin{figure*}[htbp]
    \centering
    \includegraphics[width=0.55\textwidth]{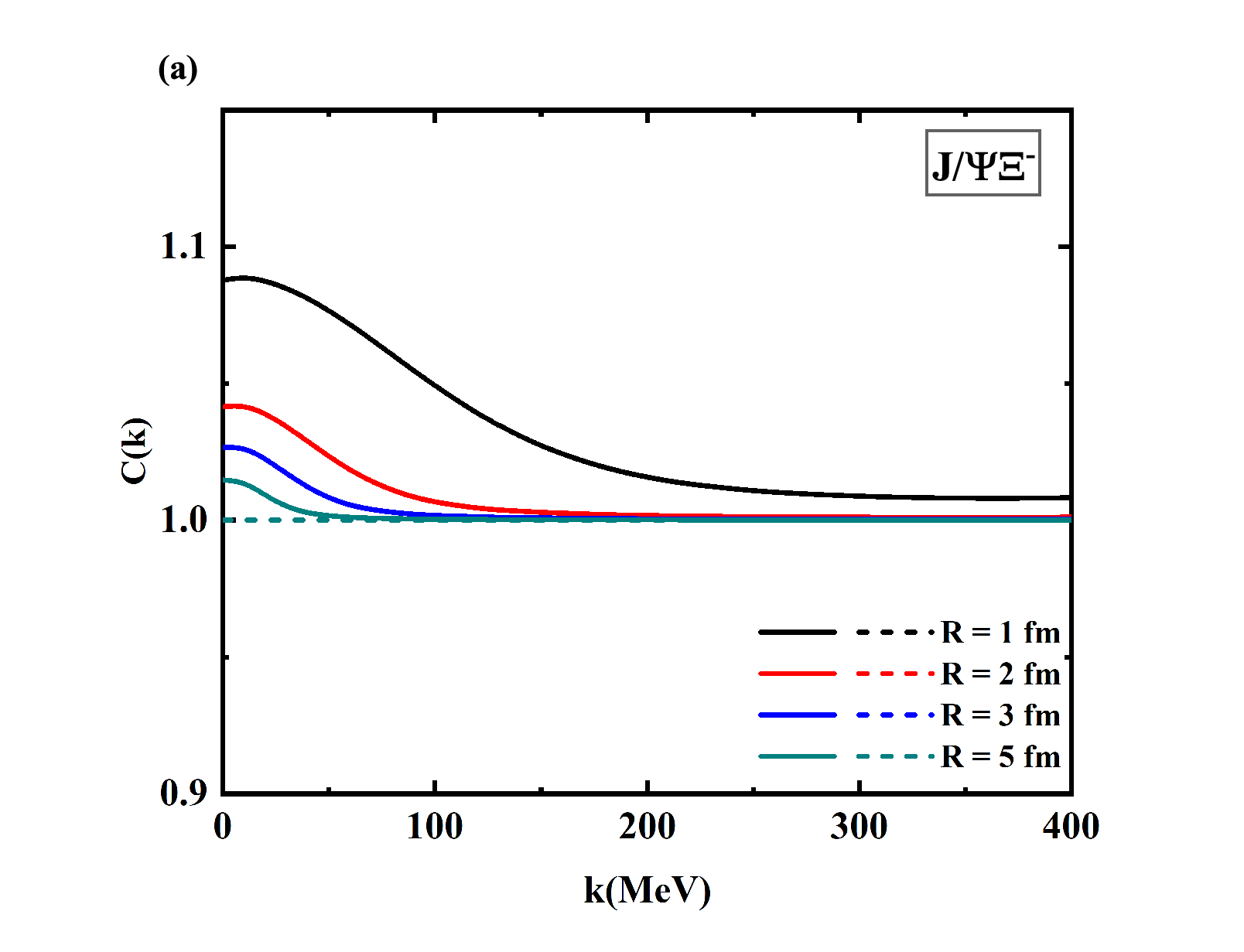}%
    \hspace{-0.1\textwidth}%
    \includegraphics[width=0.55\textwidth]{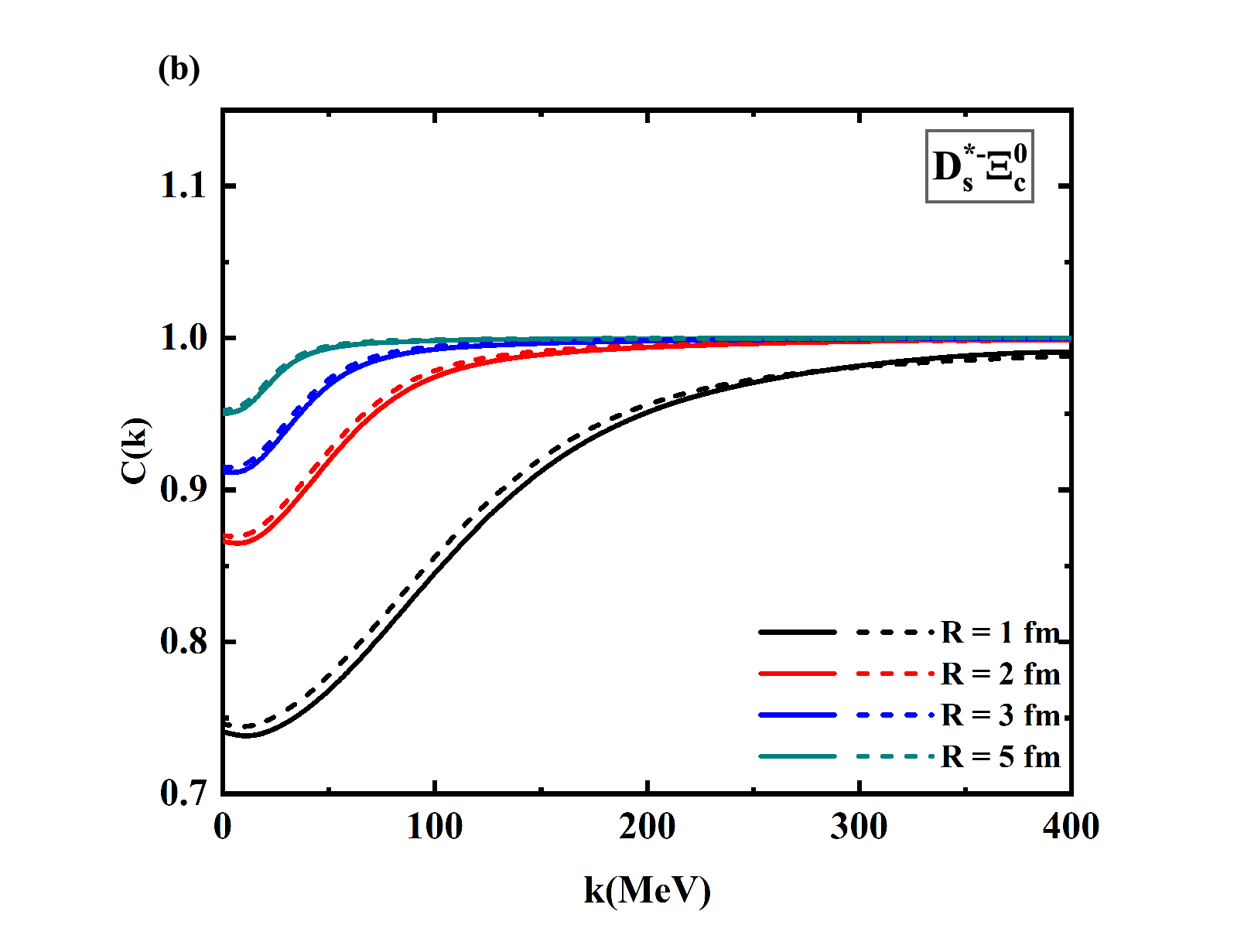}\\[-25pt]
    \includegraphics[width=0.55\textwidth]{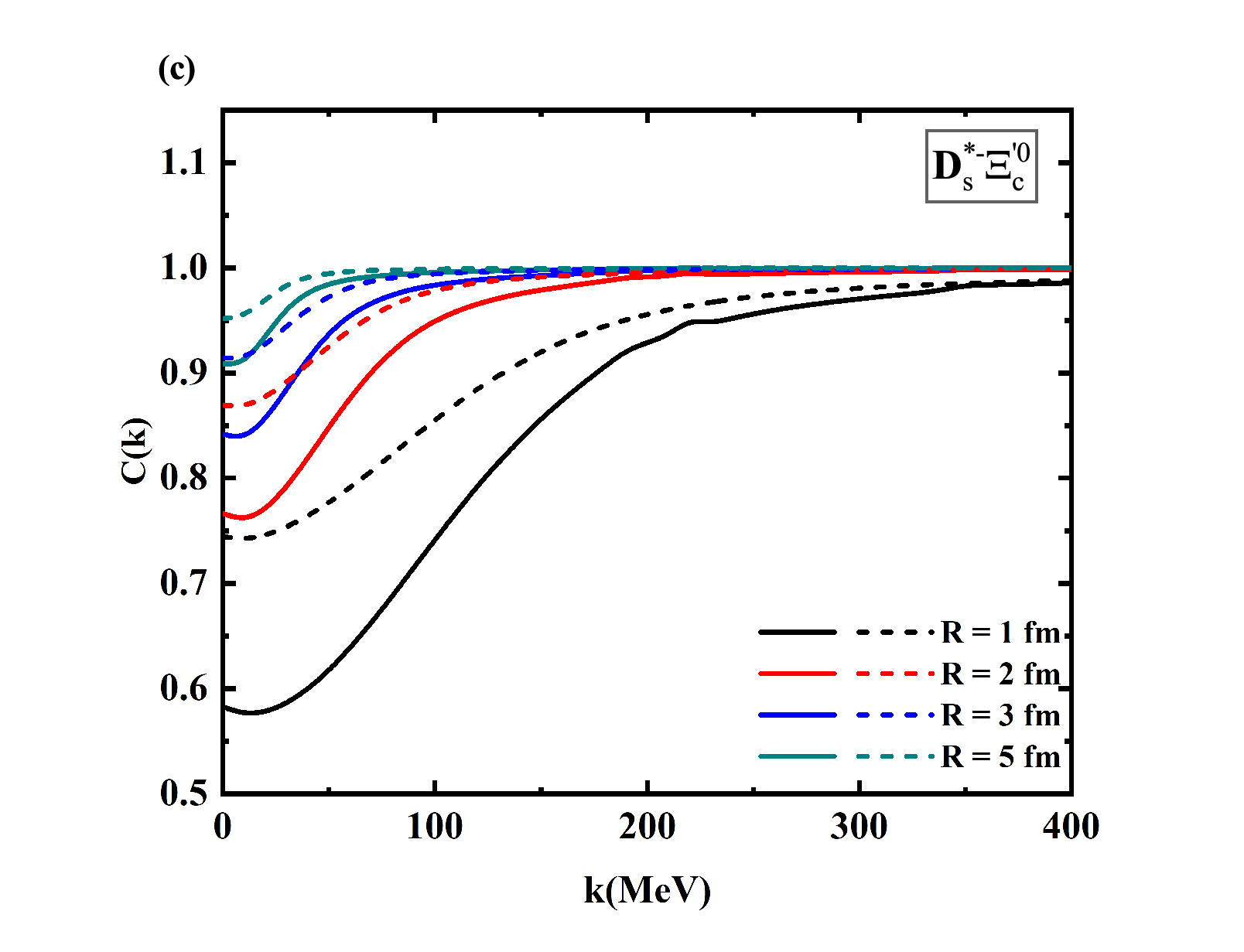}%
    \hspace{-0.1\textwidth}%
    \includegraphics[width=0.55\textwidth]{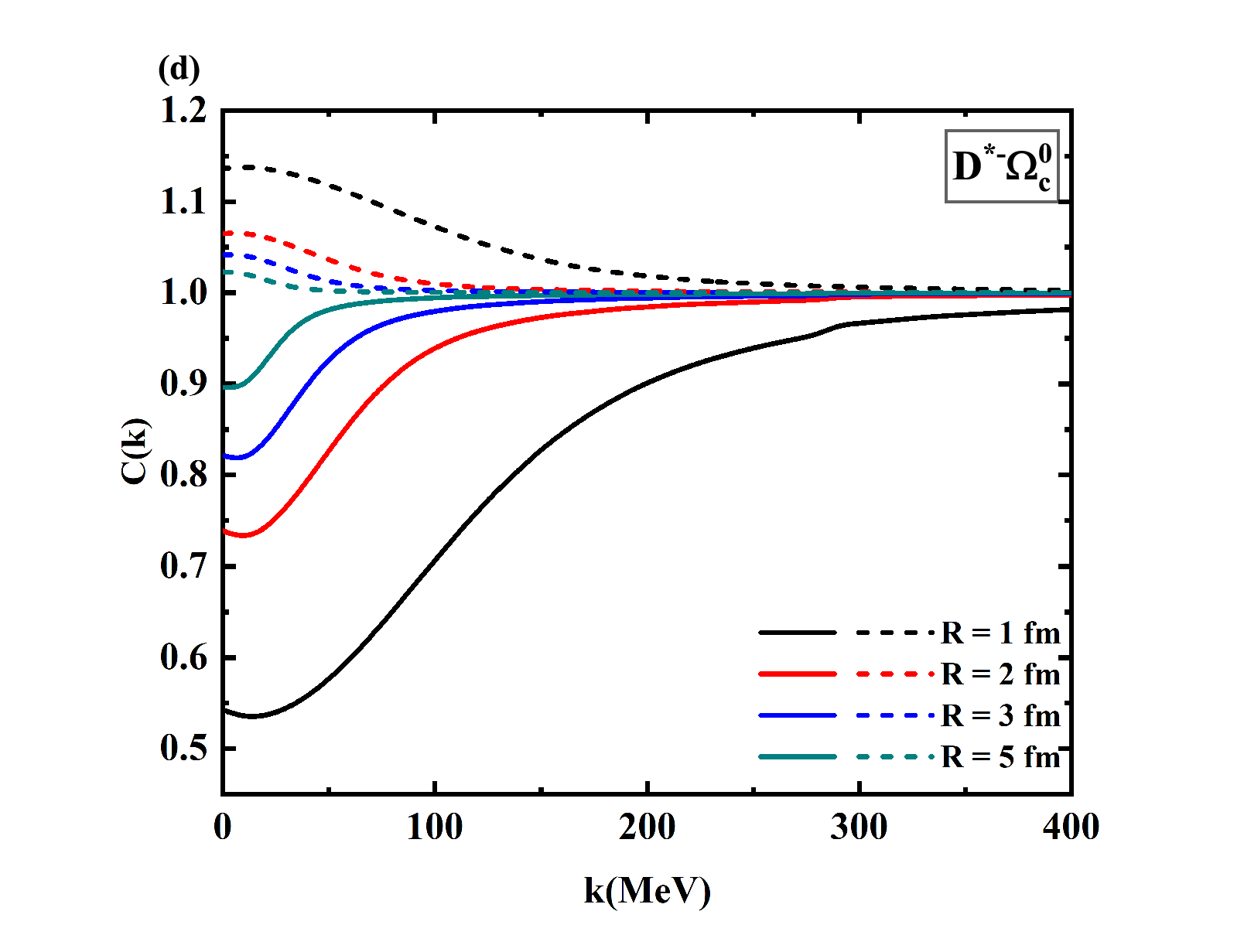}
    \caption{Same as Fig.~\ref{CF_pcss_PB} but for the $J/\psi \Xi^-$ (a), $D_s^{\ast-}\Xi_c^0$ (b), $D_s^{\ast-}\Xi_c^{\prime 0}$ (c), and $D^{\ast-}\Omega_c^0$ (d) correlation functions.}
    \label{CF_pcss_VB}
\end{figure*}

\begin{figure*}[htbp]
    \centering
    \includegraphics[width=0.55\textwidth]{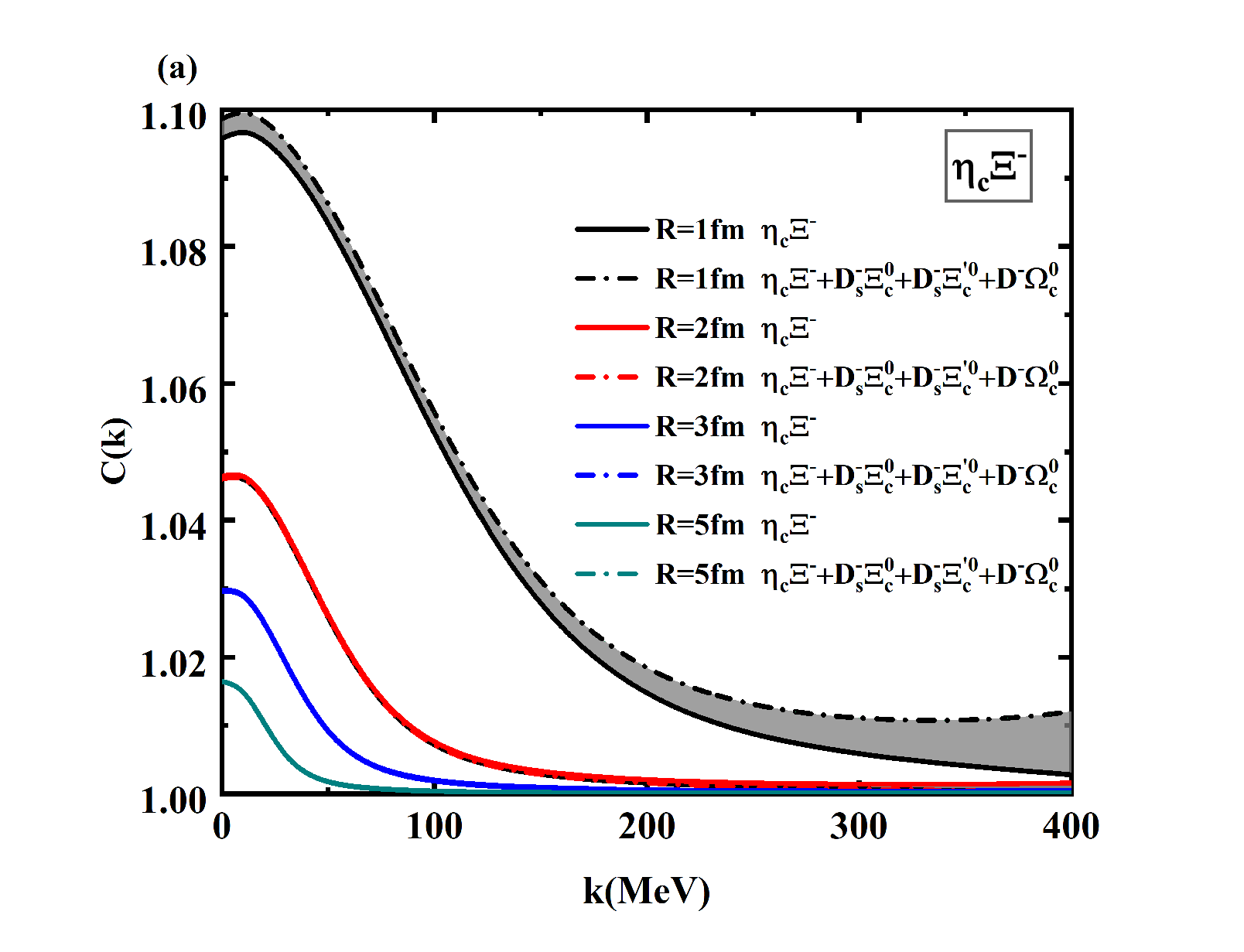}%
    \hspace{-0.1\textwidth}%
    \includegraphics[width=0.55\textwidth]{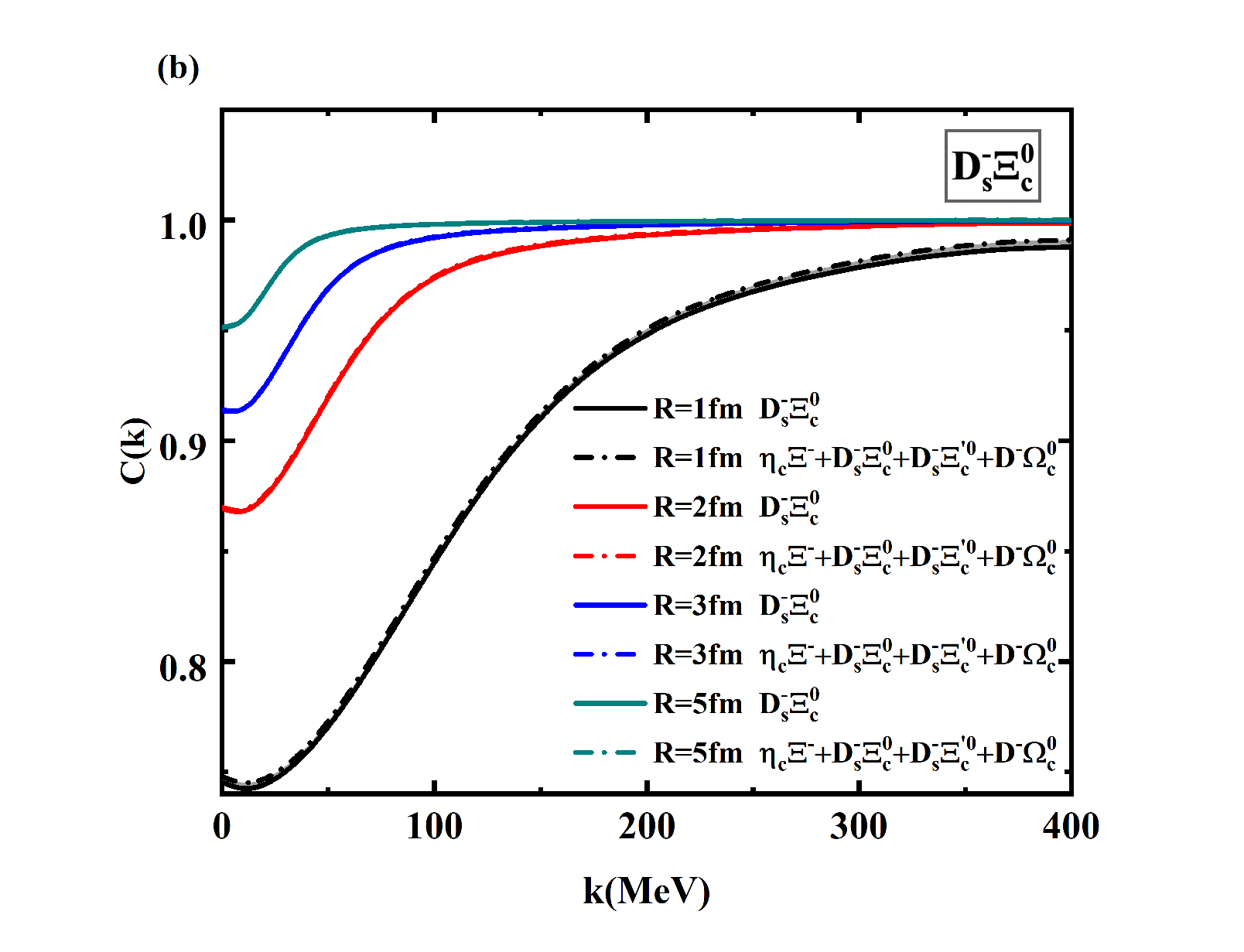}\\[-25pt]
    \includegraphics[width=0.55\textwidth]{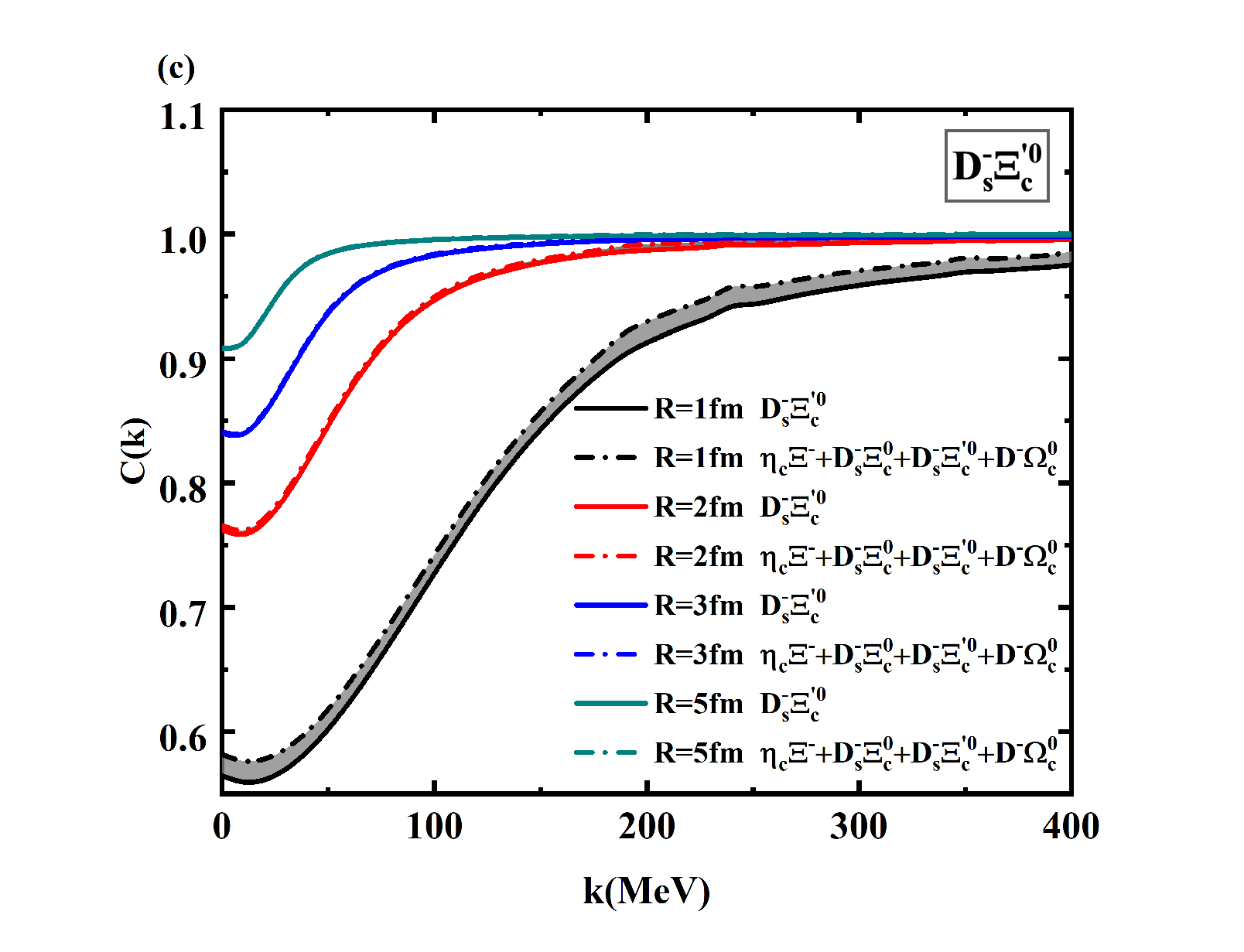}%
    \hspace{-0.1\textwidth}%
    \includegraphics[width=0.55\textwidth]{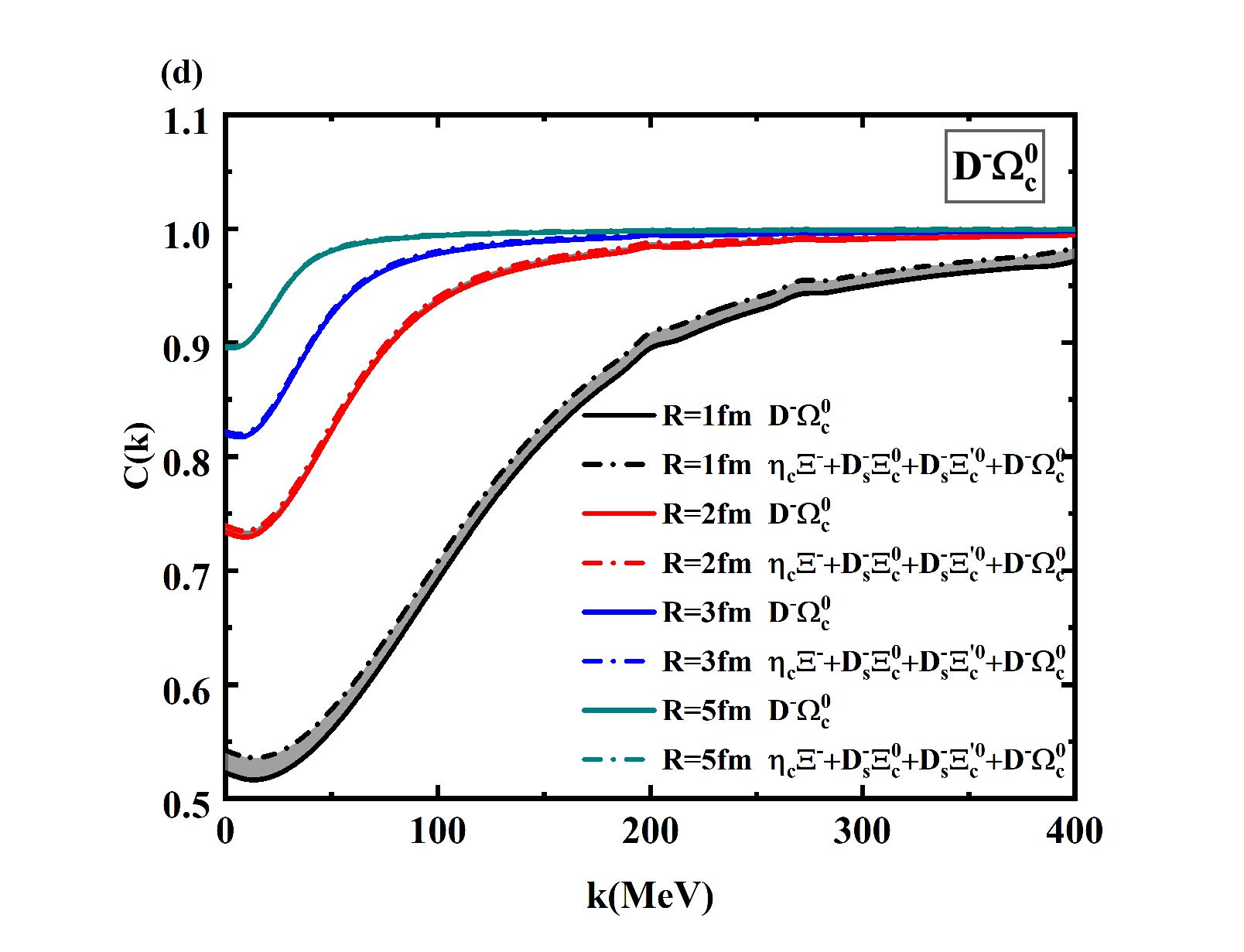}
    \caption{$\eta_{c} \Xi^-$ (a), $D_s^- \Xi_{c}^0$ (b), $D_s^- \Xi_{c}^{\prime0}$ (c), and $D^- \Omega_{c}^0$ (d) correlation functions as a function of the relative momentum $k$ for source sizes $R=1$, 2, 3, and 5 fm. The solid lines represent the results obtained in the coupled-channel case, while the short dash dot lines represent the results including only the diagonal contribution $T_{i i} \tilde{G}_{i}$. The gray shaded area represents the overlap between the full coupled-channel calculation and the calculation including only the diagonal term $T_{ii}\tilde{G}_i$, indicating that the contribution from the off-diagonal elements $T_{ji}\tilde{G}_j$ to the correlation function is negligible.}
    \label{CF_pcss_PB_CC}
\end{figure*}

\section{Results and discussions}

\subsection{Coupled-channel effects in the correlation functions}

Coupled-channel effects enter momentum correlation functions through two distinct mechanisms.  The first mechanism originates from the non-perturbative structure of the scattering amplitude, as illustrated in Eq.~(\ref{eq:CFs}), which includes all intermediate channels. The second mechanism arises from the off-diagonal elements of the scattering matrix, $T_{ji}\tilde{G}_j$ with $i\ne j$, which correspond to transitions between different reaction channels. Physically, this represents the contributions to the $i$th channel scattering wave function from the process in which particles in the initial state of channel $j$ undergo rescattering and convert into particles in channel $i$. 

To explore the influence of coupled-channel dynamics on hadron-hadron interactions, we calculate the momentum correlation functions for the PB and VB systems associated with the $P_{css}(4493)$ and $P_{css}(4633)$ states. By comparing results obtained with and without off-diagonal interactions, we aim to elucidate how coupled-channel effects modify the correlation functions and encode the underlying attractive or repulsive interactions. 

In Fig.~\ref{CF_pcss_PB}, we present the correlation functions for $\eta_{c} \Xi^-$,  $D_s^- \Xi_{c}^0$, $D_s^- \Xi_{c}^{\prime0}$, and $D^- \Omega_{c}^0$ for four typical representative source sizes $R=1$, 2, 3, and 5 fm. For comparison, we also show the results obtained in the uncoupled-channel case, in which all off-diagonal elements in Table~\ref{tab:Cij_PB} are set to zero, denoted by dashed lines. According to the general analysis in Ref.~\cite{Liu:2023uly}, the correlation functions of $D_s^- \Xi_{c}^{\prime0}$ and $D^- \Omega_{c}^0$ channels exhibit clear signatures of strong attractive interactions when taking into account full coupled-channel effects, which is consistent with the existence of the $P_{css}(4493)$ below the thresholds of both channels. 

Contrary to the aforementioned physical picture, the $D^- \Omega_c^{0}\rightarrow D^- \Omega_c^{0}$ interaction is strongly suppressed, since it proceeds only via heavy $J/\psi$ exchange. The $D_s^- \Xi_c^{\prime 0}\rightarrow D_s^- \Xi_c^{\prime 0}$ interaction is even repulsive due to the exchange of the $\phi$ meson. Thus, without considering coupled channels, one expects the correlation functions of $D^- \Omega_c^{0}$ to exhibit a totally different line shape as shown in Fig.~\ref{CF_pcss_PB} with a dashed line. The $D^- \Omega_{c}^0$ correlation function in the uncoupled-channel calculation is enhanced at low momenta, exhibiting a weakly attractive behavior. For the $D_s^- \Xi_c^{\prime 0}$ channel, on the other hand, the correlation function is suppressed below unity at low momenta, especially for sources with small sizes, as shown by the short dashed lines, indicating a repulsive interaction kernel. However, the line shapes of correlation functions for deeply bound states and repulsive interactions are similar~\cite{Liu:2023uly}. Consequently, the line-shape variations in the correlation function of $D_s^- \Xi_c^{\prime 0}$ are unlikely to be as pronounced as that of $D^- \Omega_{c}^0$, regardless of whether coupled-channel effects are included. As for the $\eta_c\Xi^-$ channel, the deviation of the correlation function from unity originates entirely from coupled-channel effects, since the interaction vanishes in the uncoupled case. This contrast indicates that the scattering amplitudes receive substantial contributions from coupled-channel transitions, which dominate their attractive behavior.

It is worth noting that although the $P_{css}(4493)$ state emerges as a resonance in the $D_s^- \Xi_c^{0}$ channel, the correlation function of this channel still exhibits a repulsive pattern~\cite{Liu:2023uly}, which is suppressed in the low-momentum region and rises as momentum increases, rather than the typical resonance behavior observed in Ref.~\cite{Liu:2024nac}, which is enhanced in the low-momentum region and decreases with increasing momentum. This is mainly because  $D_s^- \Xi_c^{0}$ cannot directly couple to the two dominant $D_s^- \Xi_{c}^{\prime0}$ and $D^- \Omega_{c}^0$ channels, as shown in Table~\ref{tab:Cij_PB}, but only indirectly via $\eta_c\Xi^-$. Such a mechanism naturally leads to a rather weak coupling of $D_s^- \Xi_c^{0}$ to $P_{css}$, and the coupled-channel effects have only a minor impact on the $T_{22}$ matrix element. The small differences between the correlation functions calculated with full coupled-channel effects and those without any coupled-channel effects confirm this picture, as shown in Fig.~\ref{CF_pcss_PB}. This suggests that the repulsive nature of the $D_s^- \Xi_c^{0}$ channel does not substantially weaken the attractive interactions in the $D_s^- \Xi_c^{\prime 0}$ and $D^- \Omega_c^{0}$ channels. To further explore the role of this channel, we increase the coupling strength between $D_s^- \Xi_c^{0}$ and $D_s^- \Xi_c^{\prime 0}$ to approximately 75\% of that between the $D_s^- \Xi_c^{\prime 0}$ and $D^- \Omega_c^{0}$ channels. Under this assumption, the correlation function of $D_s^- \Xi_c^{0}$ exhibits typical behavior in which a resonance emerges, consistent with the findings of Ref.~\cite{Liu:2024nac}. This tells that $D_s^- \Xi_c^{0}$ is not an optimal channel to study the $P_{css}(4493)$ state. A similar pattern is also observed in the VB systems. Finally, we note that the source size $R$ is selected arbitrarily for illustration. The coupled-channel effects discussed above are valid for any reasonable $R$ and are not sensitive to the specific choice of the source size $R$.

In Fig.~\ref{CF_pcss_VB}, we show the correlation functions of the vector-meson-baryon counterparts, $J/\psi \Xi^-$, $D_s^{\ast-}\Xi_c^0$, $D_s^{\ast-}\Xi_c^{\prime 0}$, and $D^{\ast-}\Omega_c^0$ pairs related to the $P_{css}(4633)$ state. Given that the spin dependence is not resolved in the standard measurements of correlation functions, we have to take the spin average for the VB interactions with $J^P=1/2^-$ and $3/2^-$. However, in the present calculation, because the potentials and the momentum cutoff are identical, the resulting correlation functions for $J^P=1/2^-$ and $3/2^-$ are identical. The solid curves obtained from the coupled-channel calculation indicate strong attractive interactions in the $D_s^{\ast-}\Xi_c^{\prime 0}$ and $D^{\ast-}\Omega_c^0$ channels. In contrast, in the $D_s^{\ast-}\Xi_c^{\prime 0}$ channel the dashed curve, which reflects the behavior of the diagonal interactions alone, exhibits a suppression below unity, indicating a repulsive interaction, while in the $D^{\ast-}\Omega_c^0$ channel, the dashed curve exhibits only a mild enhancement, corresponding to a weakly attractive interaction~\cite{Liu:2023uly}. These features are very similar to those in the PB system and support the interpretation that the strong enhancement represented by the solid curves does not originate from the diagonal terms, but rather originates predominantly from the coupled-channel dynamics encoded in $V^{\rm eff}_1$ of Eq.~(\ref{eq:2chex}).

\begin{figure}[htbp]
    \centering
    \includegraphics[width=0.5\textwidth]{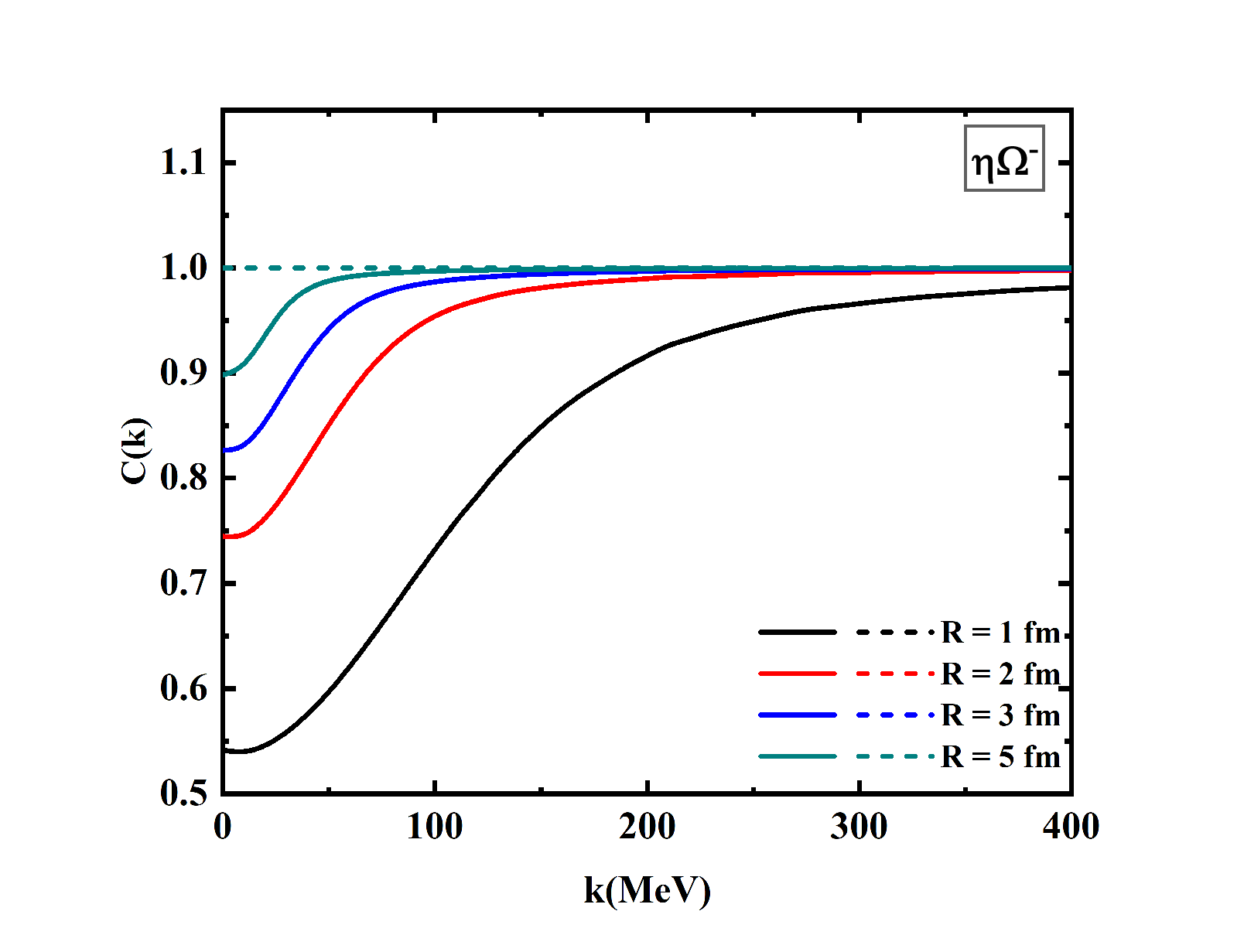}
    \caption{Same as Fig.~\ref{CF_pcss_PB} but for $\eta \Omega^{-}$ correlation function.}
    \label{etaomegacorrelationfunction}
\end{figure}

In the PB and VB interactions considered here, we find that the $T_{ji}\tilde{G}_j$ are significantly smaller than $T_{ii}\tilde{G}_i$ in the calculation of the correlation function of channel $i$. Especially, given a relatively small relative momentum $k$ of channel $i$ within our energy range of interest, e.g., $k\in [0,300]$MeV, the ratio of the loop function of channel $j$ (which has a higher threshold) to that of channel $i$, i.e., $\tilde{G}_j(r,\sqrt{s})/\tilde{G}_i(r,\sqrt{s})$ is much less than unity, due to the spherical Bessel function $j_0(kr)$. Moreover, the ratio drops sharply as the threshold difference between channels $j$ and $i$ increases.  Meanwhile, for those channels with lower thresholds compared to channel $i$, the off-diagonal element of the $T$ matrix is smaller than the diagonal $T_{ii}$ in the present system. As a consequence, the contribution from the off-diagonal element, $T_{ji}\tilde{G}_j$, to the correlation function is negligible compared with the diagonal elements. We confirm this conclusion in Fig.~\ref{CF_pcss_PB_CC} by calculating the respective contribution from the diagonal and off-diagonal $T$ matrix elements.

In this situation, we are safe to omit the last term in Eq.~(\ref{eq:CFs}) and the correlation function for channel $i$ can be simplified as
\begin{align}
   C_i(k)=&1+\int_0^\infty d r~4\pi r^2S_{12}(r)  \nonumber\\
   &\left[ |j_0(kr)+T_{ii}(\sqrt{s})\cdot \tilde{G}_i(r,\sqrt{s})|^2 - |j_0(kr)|^2 \right] \nonumber  .
\end{align}

It is even more interesting that such a conclusion is not only valid for the present PB or VB interactions. This conclusion, though maybe not universal, has also been verified in studies of the correlation function for various systems, including the $DK-D_s\eta$ system associated with $D^*_{s0}(2317)$~\cite{Liu:2023uly} and the $\bar{K}\Xi(1530)-\eta\Omega-\bar{K}\Xi$ system associated with $\Omega(2012)$~\cite{Lin:2026ypf}. Note that the contributions from coupled channels belonging to isospin multiplets cannot be neglected, since their thresholds are compatible.

The $\Omega(2012)$ provides another prominent example of a resonance dynamically generated by coupled-channel effects~\cite{Valderrama:2018bmv,Lin:2018nqd,Huang:2018wth,Pavao:2018xub,Polyakov:2018mow,Lin:2019tex,Ikeno:2020vqv,Hu:2022pae,Lu:2020ste,Ikeno:2022jpe,Ikeno:2023wyh,Lu:2022puv,Zeng:2020och,Song:2024ejc,Shen:2025xcq,Xie:2024wbd}. In the molecular picture involving $\bar{K}\Xi(1530)$ and $\eta \Omega$, all diagonal elements vanish at leading order, i.e., $V_{\bar{K}\Xi(1530) \to \bar{K}\Xi(1530)}=V_{\eta \Omega \to \eta \Omega}=0$ within the chiral perturbation theory. This implies that the $\Omega(2012)$ is solely dynamically generated by the off-diagonal interaction. The behavior of the correlation function for the $\eta \Omega^{-}$ channel relevant to $\Omega(2012)$ is shown in Fig.~\ref{etaomegacorrelationfunction}. The direct interaction of $\eta \Omega^{-}\rightarrow\eta \Omega^{-}$ vanishes at leading order. As a result, the correlation function is exactly unity in the uncoupled-channel calculation, while the correlation functions considering off-diagonal effects again exhibit a typical feature of a strongly attractive interaction. This feature clearly indicates that the deviation of the $\eta \Omega^{-}$ correlation function from unity arises solely from off-diagonal coupled-channel interactions, similar to the correlation functions of the $\eta_{c} \Xi^-$ channel in Fig.~\ref{CF_pcss_PB} and $J/\psi \Xi^-$ channel in Fig.~\ref{CF_pcss_VB}. We note that the correlation functions related to the $\Omega(2012)$ have been recently investigated in Refs.~\cite{Lin:2026ypf,Liu:2026zlk}, yielding qualitatively similar conclusions to those of the present work.

\subsection{Two-channel illustration of coupled-channel dynamics}

One notices that the correlation functions for the $D_s^- \Xi_{c}^{\prime0}$ and $D^- \Omega_{c}^0$ channels in the PB sector, as well as $D_s^{\ast-}\Xi_c^{\prime 0}$ and $D^{\ast-}\Omega_c^0$ channels in the VB sector are comparable in magnitude when coupled-channel effects are included in the calculations.

To gain further insight into how coupled-channel dynamics manifests itself in momentum correlation functions, it is useful to consider the simplest two-channel system. We assume that the diagonal interactions vanish, while only the off-diagonal transition potential survives, i.e., 
\begin{align}\label{eq:2chex}
V = 
\begin{pmatrix} 
0 & V' \\
V’ & 0 
\end{pmatrix},&~~~G = 
\begin{pmatrix} 
G_1 & 0 \\
0 & G_2 
\end{pmatrix},
\end{align}
with $V'$ the off-diagonal elements of the potential. 

From the perspective of channel 1, the effect of channel 2 can be absorbed into an effective interaction. Physically, the particles first make a transition from channel 1 to channel 2 through the off-diagonal potential $V'$, propagate in channel 2, and then return to channel 1 through another transition. This process generates the effective interaction
\begin{align}
V^{\rm eff}_1 = V' G_2 V',
\end{align}
where $G_2$ is the loop function of channel 2. Correspondingly, the scattering amplitude reads
\begin{align}
T_{11}=\frac{V^{\rm eff}_1}{1-V^{\rm eff}_1 G_1}.
\end{align}

Remarkably, a bound state or resonance may emerge in the absence of any direct interaction, given the condition
\[
1 - G_1 V^{\rm eff}_1 = 0,
\]
which is entirely governed by the off-diagonal transitions.

Now the correlation function for channel $1$ can be further simplified as
\begin{align}
\label{11}
  C_1(k)=&1+\frac{e^{-4k^2R^2}-1}{4k^2R^2} \nonumber \\
  +&\int_0^\infty d r~4\pi r^2S_{12}(r) \Bigg|j_0(kr)-\mathcal{F}(r,k)+ \nonumber\\
  &\mathcal{F}(r,k)\frac{1}{1-G_1(k) V^{\rm eff}_1(k)}\Bigg|^2 ,
\end{align}
where 
\begin{align}
\mathcal{F}(r,k)= \tilde{G}_1(r,k)/G_1(k).
\end{align}
The correlation function of channel $2$ can be obtained similarly. Notably, in this simplest two-channel situation, the only difference between the correlation functions of channels $1$ and $2$ lies in the particle masses in $\mathcal{F}(r,\sqrt{s})$ of Eq.~(\ref{11}). Particularly, under the non-relativistic expansion of the loop functions, which is equivalent to solving the Lippmann-Schwinger equation instead of the Bethe-Salpeter equation, the ratio $\mathcal{F}(r,k)$ is independent of particle masses. In the present situation involving heavy mesons and baryons, such a difference is relatively small, and it is anticipated that the correlation functions of these two channels will be very similar not only in line shapes but also in magnitude. 

The comparable correlation functions for the $D_s^{(*)-} \Xi_{c}^{\prime0}$ and $D^{(*)-} \Omega_{c}^0$ indicates the approximate validity of Eq.~(\ref{11}). Such a similarity further implies that the off-diagonal elements dominate the dynamics, while the diagonal elements play only a minor role, given the markedly different behavior in the uncoupled-channel calculations. Thus, if future experimental measurements confirm that both channels exhibit strong attraction, this would support the physical picture in which the $P_{css}(4493)$ and $P_{css}(4633)$ states are molecular states dynamically generated by the attractive interaction arising from coupled-channel effects. Specifically, the magnitude of the difference between the correlation functions of these two channels directly reveals the strength of the coupled-channel dynamics. This is a bit difficult for $D_s^{(*)-} \Xi_{c}^{\prime0}$ due to the similar behaviors of the correlation functions corresponding to the presence of a bound state and repulsive interactions.

\section{Summary}
In this work, we explored the capability of femtoscopy to probe coupled-channel effects in hadron-hadron interactions and unravel the intrinsic nature of relevant hadronic molecules. To this end, we revisited interactions of heavy pseudoscalar mesons and baryons, including $\eta_c \Xi^-$, $D_s^- \Xi_c^0$, $D_s^- \Xi_c^{\prime 0}$, and $D^- \Omega_c^0$, associated with the double strangeness pentaquark state $P_{css}(4493)$, which is governed by vector meson exchanges. Using the Koonin-Pratt formula with a Gaussian source and a unitarized coupled-channel approach, we calculated the correlation functions for all coupled channels. We further extended our study to heavy vector meson-baryon interactions and the associated $P_{css}(4633)$ state. To reveal the coupled-channel effects, we also computed the correlation functions in the uncoupled-channel case, in which all off-diagonal elements are set to zero.

In the heavy pseudoscalar meson-baryon interactions, our calculations revealed that the correlation functions of the $D_s^- \Xi_{c}^{\prime0}$ and $D^- \Omega_{c}^0$ channels display typical features of strong attractive interactions, even though their diagonal potentials are either suppressed or even repulsive. This indicates that the nature of the state is driven predominantly by the coupled-channel mechanism. Meanwhile, the correlation function of $\eta_c\Xi^-$ manifests weak attractive behavior, consistent with the nature of an above-threshold resonance~\cite{Liu:2024nac}. Notably, the elastic interaction of $\eta_c\Xi^-$ vanishes entirely, so such behavior arises purely from coupled-channel effects. In contrast, since the $D_s^- \Xi_{c}^{0}$ channel is very weakly coupled to the two dominant $D_s^- \Xi_{c}^{\prime0}$ and $D^- \Omega_{c}^0$  channels, its correlation function is almost unaffected whether coupled-channel effects are incorporated or not. A similar scenario is also found in the heavy-flavor vector meson-baryon interactions relevant to the $P_{css}(4633)$ state.

We notice that a similar physical picture exists in the interactions between decuplet baryons and pseudoscalar mesons, i.e., $\bar{K}\Xi(1530)$ and $\eta \Omega^{-}$, related to the $\Omega(2012)$, where only off-diagonal elements survive at leading order. The $\eta \Omega^{-}$ correlation function exhibits the characteristic behavior of a strong attraction, even though the leading-order diagonal interaction vanishes.

These results demonstrate that future femtoscopic measurements of these correlation functions will serve as a sensitive probe of coupled-channel dynamics, providing direct insight into the mechanisms underlying the generation of these states.

\section{Acknowledgement}
L.S.G. thanks Angel Ramos for useful communications regarding the details of Ref.~\cite{Marse-Valera:2022khy}. This work is partly supported by the National Key R\&D Program of China under Grant No. 2023YFA16067003 and the National Science Foundation of China under Grants Nos.~W2543006 and 12435007. Zhi-Wei Liu acknowledges support from the National Natural Science Foundation of China under Grant No.12522505. Zhi-Wei Liu acknowledges support from the National Natural Science Foundation of China under Grant No.12405133, and the Shenzhen Science and Technology Program under Grant No.ZDSYS20230626091501002. 
K.P.K. and A.M.T. thank the support received from the Brazilian funding agency CNPq (through Grant Nos. 407437/ 2023-1, 306461/2023-4, and 304510/2023-8).
Zheng-Ting Lai acknowledges support from Beijing Natural Science Foundation under Grant No.QY26113.

\bibliography{CCinCF}

\clearpage
\end{document}